\documentclass[prb,showpacs,preprintnumbers,amsmath,amssymb,twocolumn,superscriptaddress]{revtex4}

\usepackage{graphicx}
\usepackage{dcolumn}
\usepackage{bm}
\usepackage{color}

\begin{document}
\title{Dimensional crossover in layered $f$-electron superlattices
}

\author{Yasuhiro Tada}
\email[]{tada@issp.u-tokyo.ac.jp}
\affiliation{Institute for Solid State Physics, The University of
  Tokyo, Kashiwa 277-8581, Japan}
\author{Robert Peters}
\affiliation{Department of Physics, Kyoto University, Kyoto 606-8502,
Japan}
\author{Masaki Oshikawa}
\affiliation{Institute for Solid State Physics, The University of
  Tokyo, Kashiwa 277-8581, Japan}

\newcommand{\vecc}[1]{\mbox{\boldmath $#1$}}

\begin{abstract}
Motivated by the remarkable experimental realizations of
$f$-electron superlattices,
e.g. CeIn$_3$/LaIn$_3$- and CeCoIn$_5$/YbCoIn$_5$- superlattices,
we analyze the formation of heavy electrons in
layered $f$-electron superlattices by means
of the dynamical mean field theory.
We show that the spectral function exhibits formation of heavy
electrons in the entire system below a temperature scale $T_0$.
However, in terms of transport, two different
coherence temperatures $T_x$ and $T_z$ are identified
in the in-plane- and the out-of-plane-resistivity, respectively.
Remarkably, we find $T_z < T_x \sim T_0$ due to scatterings between
different reduced Brillouin zones.
The existence of these two distinct energy scales implies a
crossover in the dimensionality of
the heavy electrons between two and three dimensions as temperature
or layer geometry is tuned.
This dimensional crossover would be
responsible for the characteristic behaviors in the
magnetic and  
superconducting properties observed in the experiments.
\end{abstract}

\pacs{Valid PACS appear here}

\maketitle

\section{introduction}
Dimensionality plays a crucial role in condensed matter physics, 
especially in systems with strong interactions.
In most systems the dimensionality is
  determined by the structure of the material and cannot be changed.
However, in order to study the effects of dimensionality, it would be
desirable to control it.
Layered structures provide such an opportunity and make it possible to observe
effects of reduced dimensionality and crossover behavior between two and
three dimensions. 
In particular,
recent successful fabrications of the layered superlattices of 
CeIn$_3$/LaIn$_3$~\cite{pap:Shishido2010} and 
CeCoIn$_5$/YbCoIn$_5$~\cite{pap:Mizukami2012,pap:Goh2012} have opened
new possibilities for investigating such phenomena in 
$f$-electron systems. 
In these systems, the $f$-electrons are present only in the
Ce layers, which are $2$-dimensional (2D).
 However, if the $f$-electrons are coupled
through the conduction electrons of La or Yb layers,
the $f$-electrons effectively become $3$-dimensional (3D).

This question is also relevant for possible 
long range order
of
the $f$-electrons.
In the CeIn$_3(n)$/LaIn$_3$(4) superlattice \cite{pap:Shishido2010},
it was reported that, when the Ce layer
thickness $n$ is reduced to $n=2$,
the N\'eel temperature $T_N$ is decreased to zero and
the resistivity $\rho_{xx}$ shows linear temperature dependence
$\rho_{xx}\sim T$.
This non-Fermi-liquid like behavior, which is also found in Cuprates
with $2$D character, is 
in contrast to the Fermi liquid like behavior, $\rho_{xx}\sim T^2$, 
found for large $n$.
This 
suggests that the Ce layers are coupled 
and exhibit $3$D
antiferromagnetism (AF) when $n$ is large, while the coupling
between the Ce layers is suppressed
for smaller $n$ and the Ce layers retain $2$D character.

Furthermore,
superconductivity (SC) is reported
for
the CeCoIn$_5(n)$/YbCoIn$_5(5)$
superlattice~\cite{pap:Mizukami2012,pap:Goh2012}.
For 
CeCoIn$_5(n)$-layer
thickness $n\geq 3$,
clear SC transitions are found.
In particular, for $n=5$, the field angle dependence of the upper critical field
$H_{c2}$ slightly below the transition temperature can be well fitted 
by the $3$D anisotropic mass model. 
In contrast, 
for $n=3$, the $3$D anisotropic mass model can no longer
explain $H_{c2}$, and $H_{c2}$ is well fitted by the Tinkham model for
thin superconductors, in which the thickness of the system is smaller than
the $z$-axis SC coherence length \cite{book:Tinkham}.
This implies that the 
superconducting Ce layers
are coupled to form
a $3$D superconductor when $n$ is large,
while the coupling is suppressed for smaller $n$ 
so that the Ce layers
remain $2$D superconductors.

Magnetism and SC in these layered $f$-electron superlattices
discussed above are interesting problems, much of which
is still open.
At the same time, 
these experimental results raise an even more
fundamental question 
of the dimensionality of the heavy electron states
due to the $f$-electrons, even in the absence of any order.
Theoretical approaches to this problem so far have been based on an
implicit assumption that the $f$-electrons, separated by the spacer
layers, are almost decoupled, 
which results
in essentially $2$D
heavy electron states~\cite{pap:Maruyama2012,pap:She2012}.
However, whether the heavy electrons in the superlattice
are actually $2$D or not is a non-trivial issue.
Understanding this issue would also be essential in
tackling questions about magnetism and SC.

In this paper, we study the formation of the heavy electrons through
the Kondo effect and their properties in layered $f$-electron superlattices.
We clarify the dimensionality of the heavy electrons in
  the superlattice
and discuss qualitatively the experimental observations of the
AF properties in CeIn$_3$/LaIn$_3$ and the
anomalous $H_{c2}$ in CeCoIn$_5$/YbCoIn$_5$ based on dimensional crossover.

\section{Model}
In order to capture the essential points of layered
  $f$-electron superlattices, we use a model which describes
a system with two kinds of layers. One kind of layers
include both $c$- and $f$-electrons
(corresponding to Ce layers),
and
the other kind of layers
include only $c$-electrons (corresponding to
La or Yb layers). 
We call the former type of layers ``A-layers'',
and the latter type of spacer layers ``B-layers''.
It is noted that, the density of states around the Fermi energy in a
CeCoIn$_5$/YbCoIn$_5$ superlattice 
is almost completely determined by the electrons on the Ce-sites 
and the Yb-sites~\cite{pap:Mizukami2012},
which validates our model for the superlattices.
The numbers of A-layers and B-layers within the unit cell 
are given by
$L_A$ and $L_B$, respectively, and $L\equiv L_A+L_B$ is
the thickness of the unit cell.
Each layer is represented by a square lattice.
The Hamiltonian, which is a variant of the Periodic Anderson Model (PAM),
reads
\begin{align}
H&=-t_c\sum_{izjz^{\prime}\sigma}c^{\dagger}_{iz\sigma}c_{jz^{\prime}\sigma}
-t_f\sum_{izjz^{\prime}\in {\rm A}, \sigma}
f^{\dagger}_{iz\sigma}f_{jz^{\prime}\sigma}\notag \\
&+V\sum_{iz\in {\rm A},\sigma}[c^{\dagger}_{iz\sigma}f_{iz\sigma}
+f^{\dagger}_{iz\sigma}c_{iz\sigma}]\notag \\
&+U\sum_{iz\in {\rm A}}[n_{iz\uparrow}^f-1/2][n_{iz\downarrow}^f-1/2],
\end{align}
where $i,j=(x,y)$ correspond to in-plane sites, 
$z$ is the layer index, and $\sigma$ is the spin index.
Hopping is only allowed between nearest neighbor sites.
We set the chemical potentials such that the particle-hole symmetry is
 conserved.
Because $t_f<V<t_c$ is satisfied for the bare parameters in many $f$-electron systems, we fix them as $t_f=0.2, V=0.4$ as a typical set of values, taking $t_c=1$ as the energy unit. We also fix $U=2.4=15V^2/t_c$, which leads to a renormalized hopping $t^{\ast}_f=t_f/(1-\partial \Sigma^{ff}(0)/\partial \omega)\sim 0.03-0.04 t_c$ in 3D PAM. For these parameters, the resistivities show pronounced peaks when the temperature is changed as observed in the experiments for bulk CeIn$_3$ and CeCoIn$_5$.
The qualitative physics described in this paper is unchanged for different
 parameters as long as $t_f<V<t_c$  and $U$ is large enough.
Note that, due to non-zero $t_f$, the system is metallic
even at half filling.
We emphasize that, our model is based on  a
standard model for $f$-electron systems, PAM,
and fully incorporates the superlattice structure.
In this respect, the present model is a minimal
microscopic Hamiltonian for $f$-electron superlattices,
including both essential ingredients.
Because the materials used for different layers in the
experimental setup are charge-neutral, we do not consider any effects of
charge redistribution in this study.

For our Hamiltonian, we have imposed periodic boundary conditions in all 
directions, so that we can perform Fourier transformation.
The Fourier transformation is given by
\begin{align}
c_{jz\sigma}&=\sum_{k_{\parallel}k_zl}
U^{c}_{j\tilde{z}_1\tilde{z}_2,k_{\parallel}k_zl}
c_{k_{\parallel}k_zl\sigma},\\
f_{jz\sigma}&=\sum_{k_{\parallel}k_zl}
U^{f}_{j\tilde{z}_1\tilde{z}_2,k_{\parallel}k_zl}
f_{k_{\parallel}k_zl\sigma},
\end{align}
where the unitary matrices $U^c$ and $U^f$ are defined as,
\begin{align}
U^c_{j\tilde{z}_1\tilde{z}_2,k_{\parallel}k_zl}
&=\frac{e^{ik_{\parallel}R_{j\parallel}}}{\sqrt{N_{\parallel}}}
\frac{e^{ik_zz+iq^c_l\tilde{z}_2}}{\sqrt{N_z}},\\
U^f_{j\tilde{z}_1\tilde{z}_2,k_{\parallel}k_zl}
&=\frac{e^{ik_{\parallel}R_{j\parallel}}}{\sqrt{N_{\parallel}}}
\frac{e^{ik_zz+iq^f_l\tilde{z}_2}}
{\sqrt{N_zL_A/L}}.
\end{align}
Here $\vecc{k}_{\parallel}=(k_x,k_y)$ 
and $\vecc{R}_{j\parallel}
=(x_j,y_j)$.
The layer index $z$ is parametrized as
$z=L\tilde{z}_1+\tilde{z}_2$ with $0\leq \tilde{z}_2<L$ for $U^c$
and $0\leq \tilde{z}_2<L_A$ for $U^f$.
The "orbital" index $l$ is $0\leq l< L (L_A)$ for $U^c (U^f)$,
and $q^c_l=2\pi l/L, q^f_l=2\pi l/L_A$.
The momentum along the $z$-axis is defined in the reduced
Brillouin zone (RBZ), $0\leq k_z<2\pi/L$.
$N_{\parallel}$ is the total number of the sites within a layer and 
$N_z$ is the total number of the layers.
The total number of sites is $N=N_{\parallel}N_z$.
We note that the orbital index $l$
arises from the superlattice structure through the 
Fourier-transformation.
Roughly speaking, it specifies 
where the momentum $K_z=k_z+q^c_l$ is located in the unfolded RBZ
$[0,2\pi)=\cup_{l=1}^{L}[2\pi(l-1)/L,2\pi l/L)$. 
We note that 
only $k_z$ is conserved while $K_z$ is 
generally not conserved.
In the new basis, the Hamiltonian becomes
\begin{align}
H&=\sum_{k\sigma} \sum_{aa^{\prime},ll^{\prime}}
A^{a\dagger}_{kl\sigma}H^{aa^{\prime}}_{ll^{\prime}}(\vecc{k})
A^{a^{\prime}}_{kl^{\prime}\sigma}\notag \\
&\quad +\frac{U}{2NL_A/L}\sum_{\{k_i,l_i\}\sigma}
f^{\dagger}_{k_1l_1\sigma}f_{k_2l_2\sigma}
f^{\dagger}_{k_3l_3\bar{\sigma}}f_{k_4l_4\bar{\sigma}}\notag \\
&\qquad \qquad \times\delta_{k_1+k_3,k_2+k_4}
\delta_{l_1+l_3,l_2+l_4},
\end{align}
where $A^{a=c}=c,A^{a=f}=f$, and $\vecc{k}=(\vecc{k}_{\parallel},k_z)$.
The elements of the Hamiltonian are
\begin{align}
H^{cc}_{ll^{\prime}}&=-2t_c[\cos k_x+\cos k_y+\cos (k_z+q^c_l)]
\delta_{ll^{\prime}}\notag \\
&\qquad \qquad \qquad (0\leq l,l^{\prime}< L),\\
H^{ff}_{ll^{\prime}}&=-2t_f[\cos k_x+\cos k_y]\delta_{ll^{\prime}}\notag \\
&\quad -t_fS^{ff}_{ll^{\prime}}[e^{-i(k_z+q^f_l)}
+e^{i(k_z+q^f_{l^{\prime}})}] \quad (0\leq l,l^{\prime}< L_A),\\
H^{cf}_{ll^{\prime}}&=VS^{cf}_{ll^{\prime}}
\quad (0\leq l< L, 0\leq l^{\prime}< L_A),\\
H^{fc}_{ll^{\prime}}&=H^{cf\ast}_{l^{\prime}l},
\end{align}
where 
\begin{align}
S^{ff}_{ll^{\prime}}&=\frac{1}{L_A}\sum_{z=0}^{z_0}
e^{-i(q^f_l-q^f_{l^{\prime}})z},\\
S^{cf}_{ll^{\prime}}&=\frac{1}{\sqrt{LL_A}}\sum_{z=0}^{L_A-1}
e^{-i(q^c_l-q^f_{l^{\prime}})z},
\end{align}
with $z_0=L_A-2$ for $L_B\neq0$ and $z_0=L_A-1$ for $L_B=0$.
We note that none of the elements of $H_{ll^{\prime}}^{cf,fc}$
vanishes, and therefore, all the $c$-electrons are
coupled to the $f$-electrons as long as $V\neq0$.
This suggests that all states on the Fermi surface at 
very low temperature
should be composite states of the $c$-electrons and the $f$-electrons.

The Green's functions in this basis are given by
\begin{align}
G^{aa^{\prime}}_{ll^{\prime}}(i\omega_n,\vecc{k})
=-\int_0^{1/T}d\tau \langle T_{\tau}A^a_{kl\sigma}(\tau)
A^{a^{\prime}\dagger}_{kl^{\prime}\sigma}(0)\rangle e^{i\omega_n\tau}.
\end{align}
We note that 
off-diagonal elements $G_{ll^{\prime}}(l\neq l^{\prime})$ do
not vanish and that these
"inter-orbital" elements include scattering processes between
different RBZs.
This point is essentially important for understanding the anisotropic
behavior in the resistivity as will be discussed in the next section.

Correlation effects 
are taken into account by means of
the inhomogeneous dynamical mean field theory
(DMFT)\cite{georges1996,pap:Potthoff1999,pap:Okamoto2004,pap:Helmes2008,pap:Zenia2009}
combined with the numerical renormalization 
group (NRG) as an impurity 
solver \cite{Wilson1975,Bulla2008,peters2006,weichselbaum2007,com:NRG}.
Although the selfenergy $\Sigma$ is site-diagonal,
$\Sigma$ differs for each A-layer $\Sigma_{izjz^{\prime}}(\omega)
=\Sigma_{z}(\omega)\delta_{ij}\delta_{zz^{\prime}}$ in this approximation.
Because DMFT+NRG appropriately takes
local correlations into account,
the formation of heavy electrons through the Kondo effect
is well described by this method.
The self-consistent equation reads,
\begin{align}
{\mathcal G}_{z}(\omega)
=\biggl[\frac{1}{N_{\parallel}}\sum_{k_{\parallel}}G_{zz}(\omega,\vecc{k}_{\parallel})\biggr]^{-1}
+\Sigma_{z}(\omega),
\end{align}
where ${\mathcal G}$ and $G$ are the cavity Green's function and 
the lattice Green's function, respectively.
The lattice Green's function $G_{zz^{\prime}}(\omega,\vecc{k}_{\parallel})$
is obtained by the inverse Fourier transformation
with respect to $k_z,l$ from $G_{ll^{\prime}}(\omega,\vecc{k})$

\section{calculation results}
As was discussed in the previous sections, dimensionality of a superlattice
in the paramagnetic normal states
is a fundamental property and is a key for understanding the experiments.
In this section, we discuss 
the dimensionality of the system based on our numerical results
within the DMFT calculations.
In order to clarify the dimensionality, we investigate two measures
of dimensionality: band
structures of the system and resistivities in different
directions.

\begin{figure}[htbp]
\begin{tabular}{cc}
\begin{minipage}{0.5\hsize}
\begin{center}
\includegraphics[width=0.95\hsize,height=0.95\hsize]{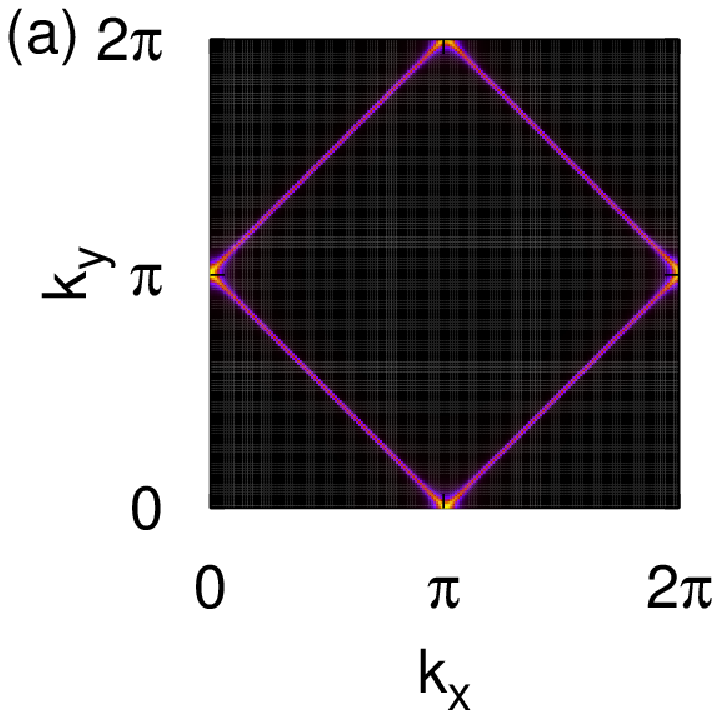}
\end{center}
\end{minipage}
\begin{minipage}{0.5\hsize}
\begin{center}
\includegraphics[width=0.95\hsize,height=0.95\hsize]{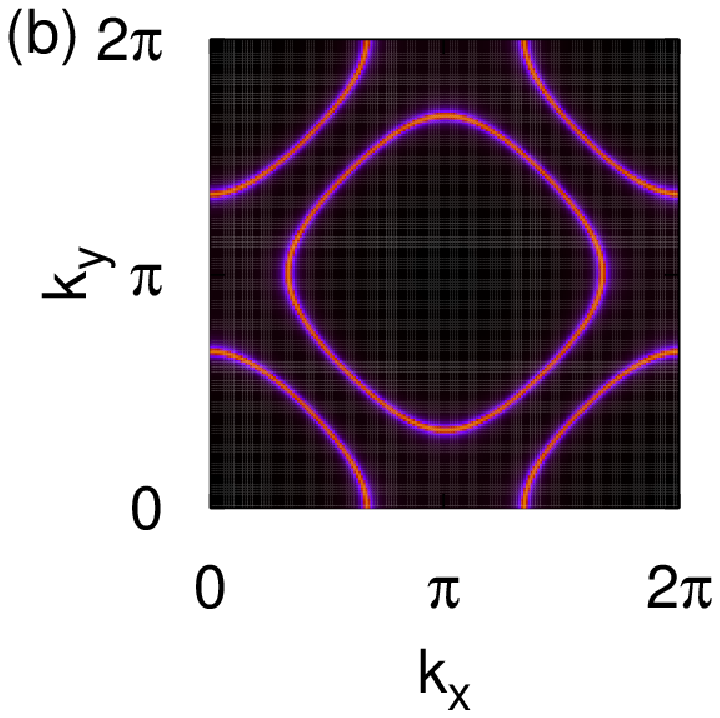}
\end{center}
\end{minipage}
\end{tabular}
\begin{tabular}{cc}
\begin{minipage}{0.5\hsize}
\begin{center}
\includegraphics[width=\hsize,height=0.95\hsize]{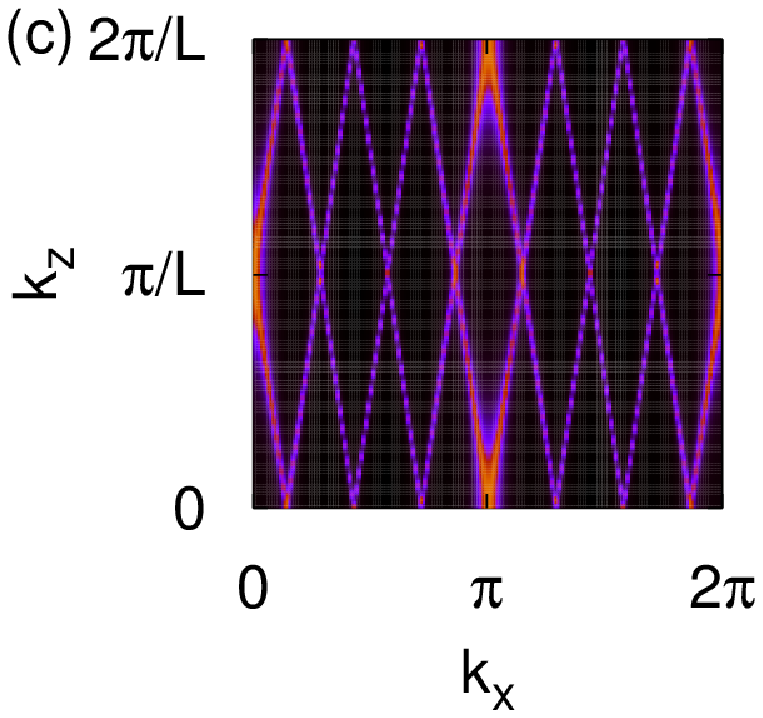}
\end{center}
\end{minipage}
\begin{minipage}{0.5\hsize}
\begin{center}
\includegraphics[width=\hsize,height=0.95\hsize]{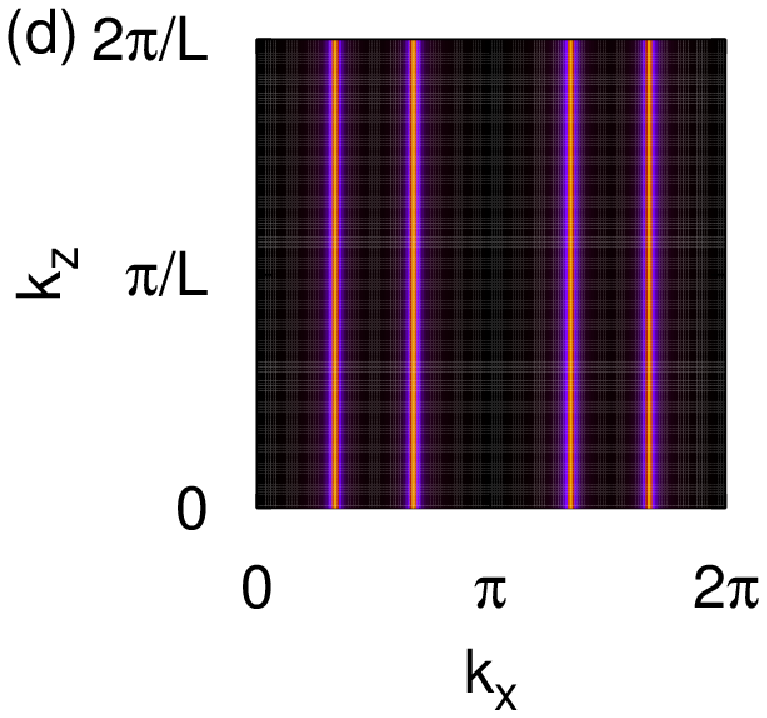}
\end{center}
\end{minipage}
\end{tabular}
\caption{(color online)
The Fermi surface with $U=0, V=0$ for
(a) $c$-electrons at $K_z=q_1^c+k_z=
\pi/2=2\pi/7+3\pi/14$, (b) $f$-electrons,
(c) $c$-electrons at $k_y=\pi/2$, and (d) $f$-electrons
at $k_y=\pi/2$.
The Fermi surfaces in (a) and (c) are the same.
Violet corresponds to high-intensity regions and black corresponds to 
low-intensity regions.
}
\label{fig:FS_V0}
\end{figure}

We start our discussion by analyzing the spectral function
\begin{align}
A(\omega,\vecc{k})=-\frac{1}{\pi}{\rm tr}\bigl[
{\rm Im}G^R(\omega,\vecc{k})\bigr], 
\end{align}
where $\vecc{k}=(k_x,k_y,k_z)$ with 
$0\leq k_{x,y}<2\pi$ and $0\leq k_z<2\pi/L$, and
$G^R$ is the retarded Green's function.
Here, we focus on a $(L_A,L_B)=(2,5)$-superlattice which is exemplary
for $f$-electron superlattices. 
\begin{figure}[htbp]
\begin{tabular}{cc}
\begin{minipage}{0.5\hsize}
\begin{center}
\includegraphics[width=0.95\hsize,height=0.95\hsize]{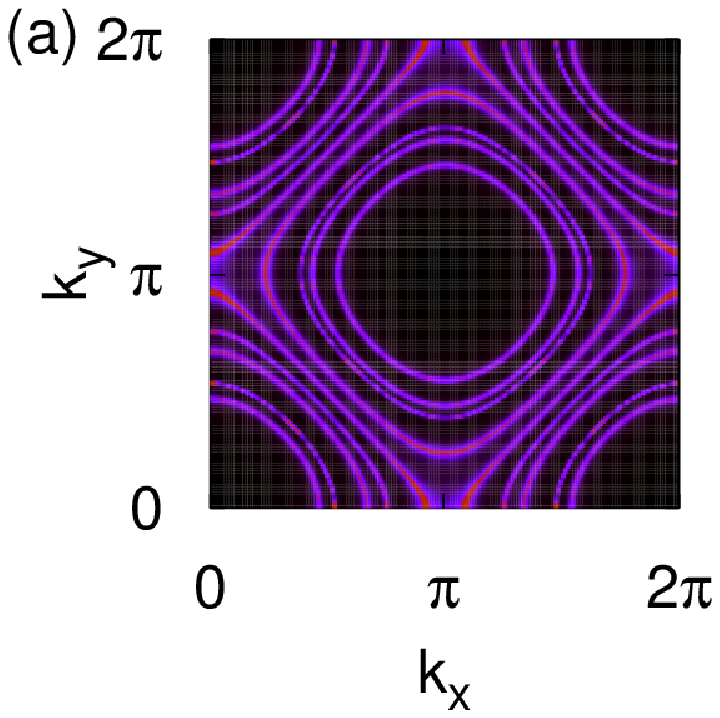}
\end{center}
\end{minipage}
\begin{minipage}{0.5\hsize}
\begin{center}
\includegraphics[width=0.95\hsize,height=0.95\hsize]{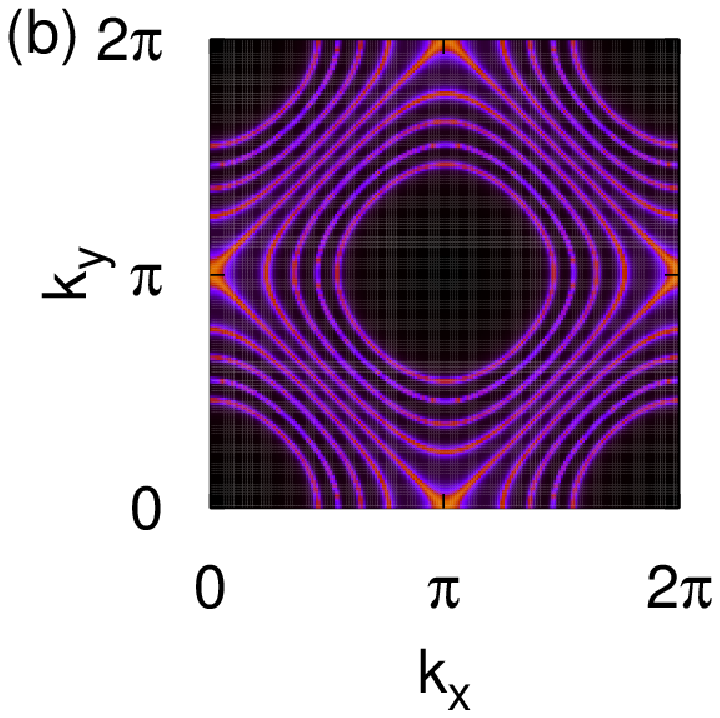}
\end{center}
\end{minipage}
\end{tabular}
\begin{tabular}{cc}
\begin{minipage}{0.5\hsize}
\begin{center}
\includegraphics[width=\hsize,height=0.95\hsize]{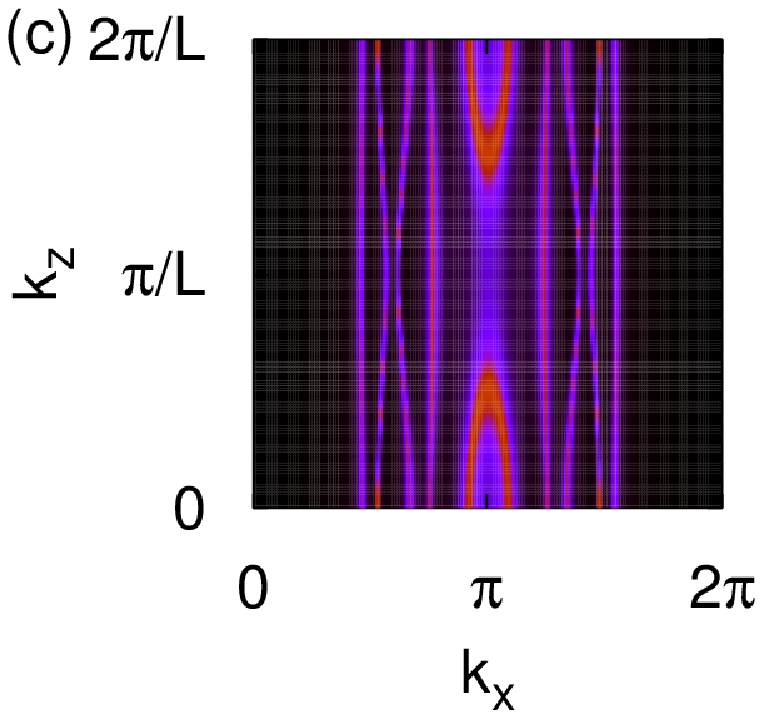}
\end{center}
\end{minipage}
\begin{minipage}{0.5\hsize}
\begin{center}
\includegraphics[width=\hsize,height=0.95\hsize]{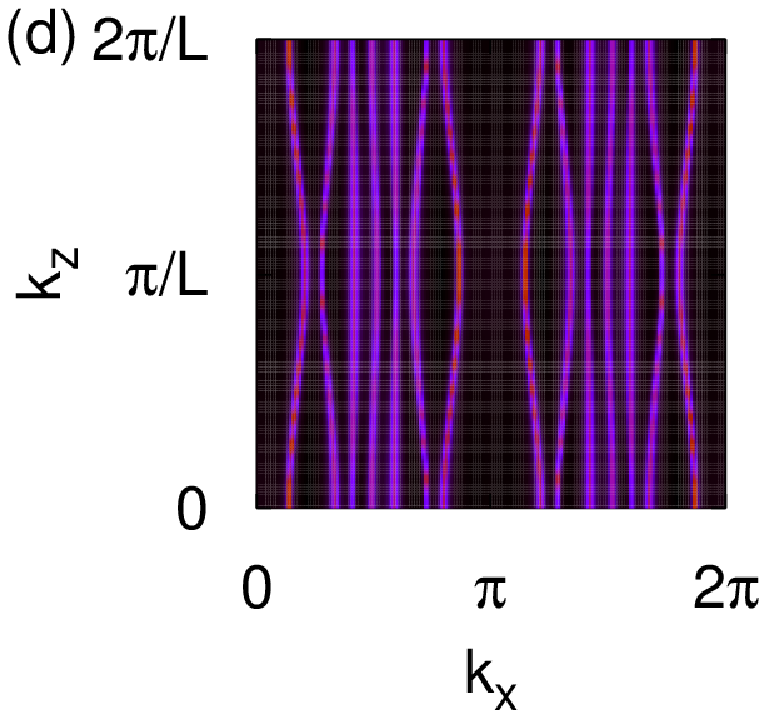}
\end{center}
\end{minipage}
\end{tabular}
\caption{(color online)
The spectral function $A(\omega=0,\vecc{k})$ for $(L_A,L_B)=(2,5)$
at (a) $k_z=0$,
(b) $k_z=\pi/2L$, (c) $k_y=0$
and (d) $k_y=\pi/2$. Tempterature is $T=0.0015$.
Violet corresponds to high-intensity regions and black corresponds to 
low-intensity regions.
}
\label{fig:FS}
\end{figure}
First, we discuss the shape of the Fermi surface in the superlattice
by looking at
$A(\omega,\vecc{k})$ for $\omega=0$.
In Fig. \ref{fig:FS}, 
we show contour plots of  $A(\omega=0,\vecc{k})$ for $(L_A,L_B)=(2,5)$
at sufficiently low temperature
$T=0.0015$. 
This temperature is much lower than the coherence temperature
so that heavy quasi-particles are well formed in these figures
as will be discussed later.
Violet corresponds to high-intensity regions and black corresponds to 
low-intensity regions.
One can compare the Fermi surface in Fig. \ref{fig:FS}
with the Fermi surface for $V=0, U=0$ in Fig. \ref{fig:FS_V0} where
the $c$-electron Fermi surface is isotropic and the $f$-electrons
have a 2-dimensional Fermi surface.
Compared to the Fermi surface for $V=0, U=0$,
the Fermi surface in the superlattice is strongly anisotropic.
However, we can clearly see that the Fermi surface in the $k_xk_z$-plane has
finite curvatures in Fig. \ref{fig:FS}. 
Therefore, we can state that
low energy quasi-particles in the superlattice have finite velocity 
in the $z$-direction.

\begin{figure}[htbp]
\begin{tabular}{cc}
\begin{minipage}{0.5\hsize}
\begin{center}
\includegraphics[width=\hsize,height=0.8\hsize]{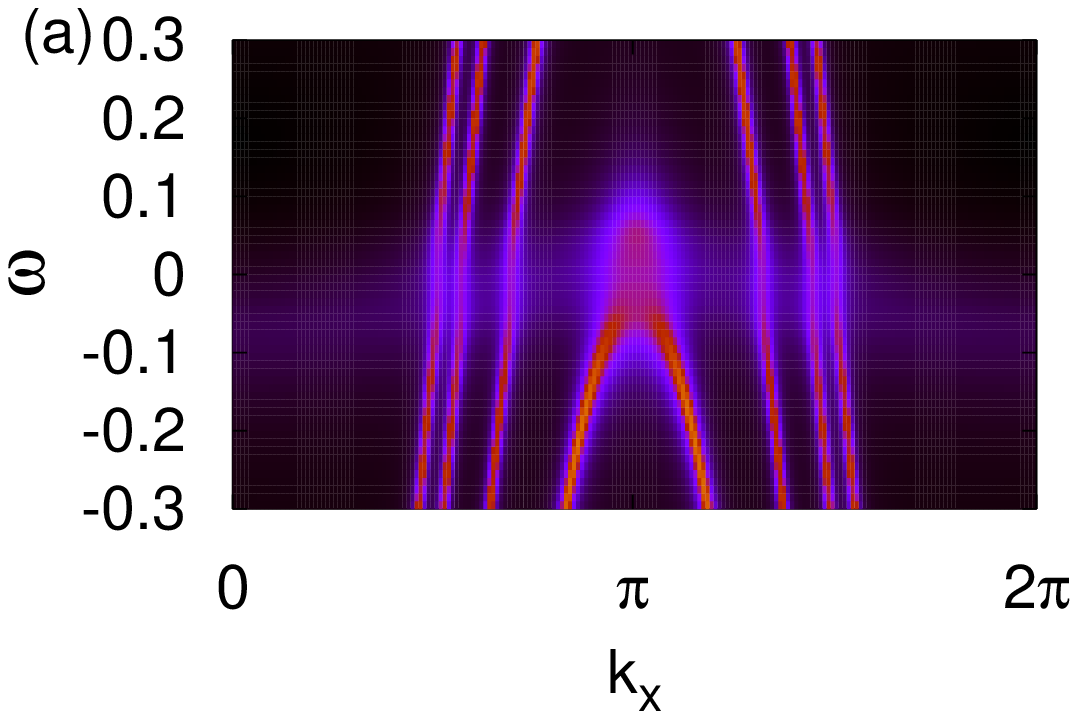}
\end{center}
\end{minipage}
\begin{minipage}{0.5\hsize}
\begin{center}
\includegraphics[width=\hsize,height=0.8\hsize]{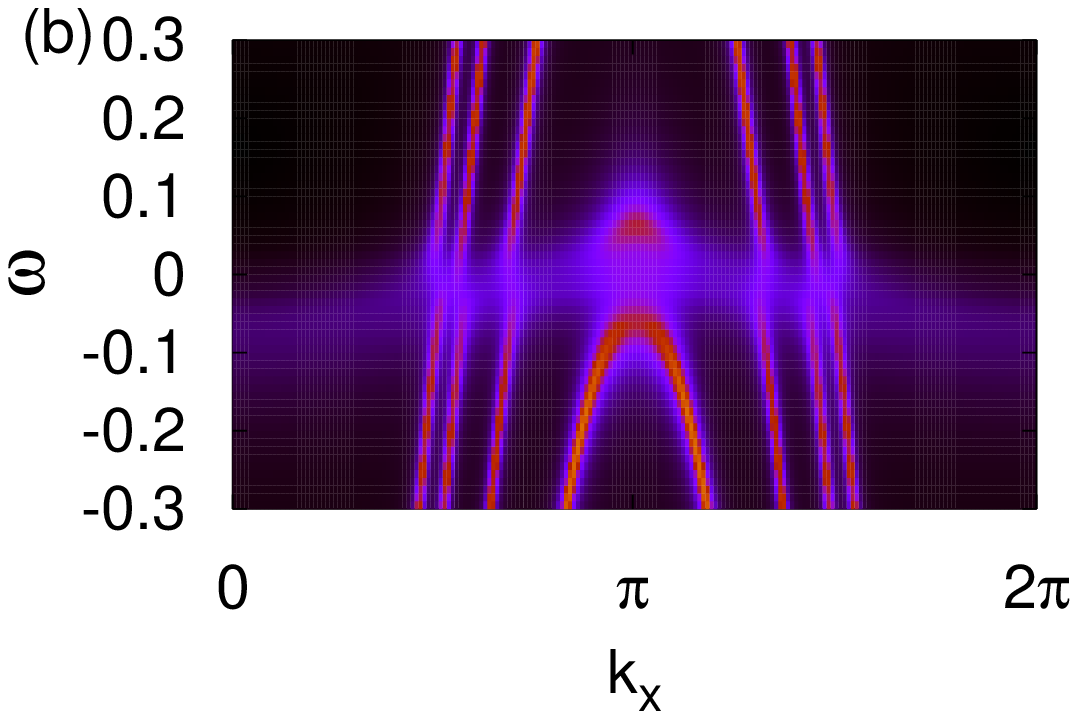}
\end{center}
\end{minipage}
\end{tabular}
\begin{tabular}{cc}
\begin{minipage}{0.5\hsize}
\begin{center}
\includegraphics[width=\hsize,height=0.8\hsize]{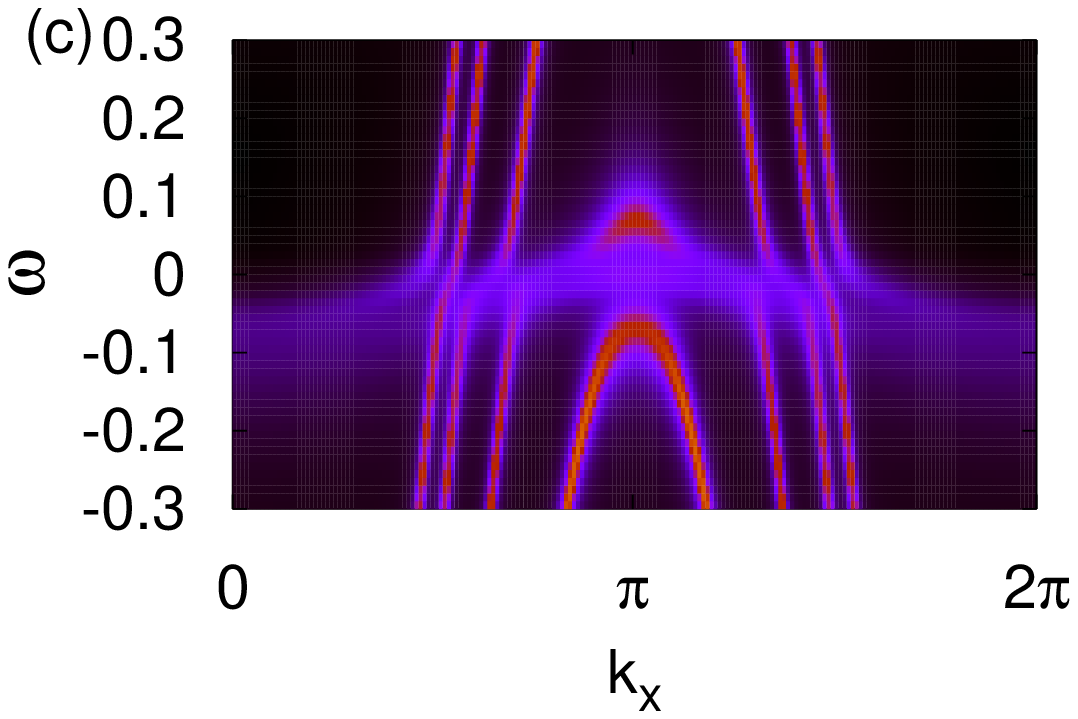}
\end{center}
\end{minipage}
\begin{minipage}{0.5\hsize}
\begin{center}
\includegraphics[width=\hsize,height=0.8\hsize]{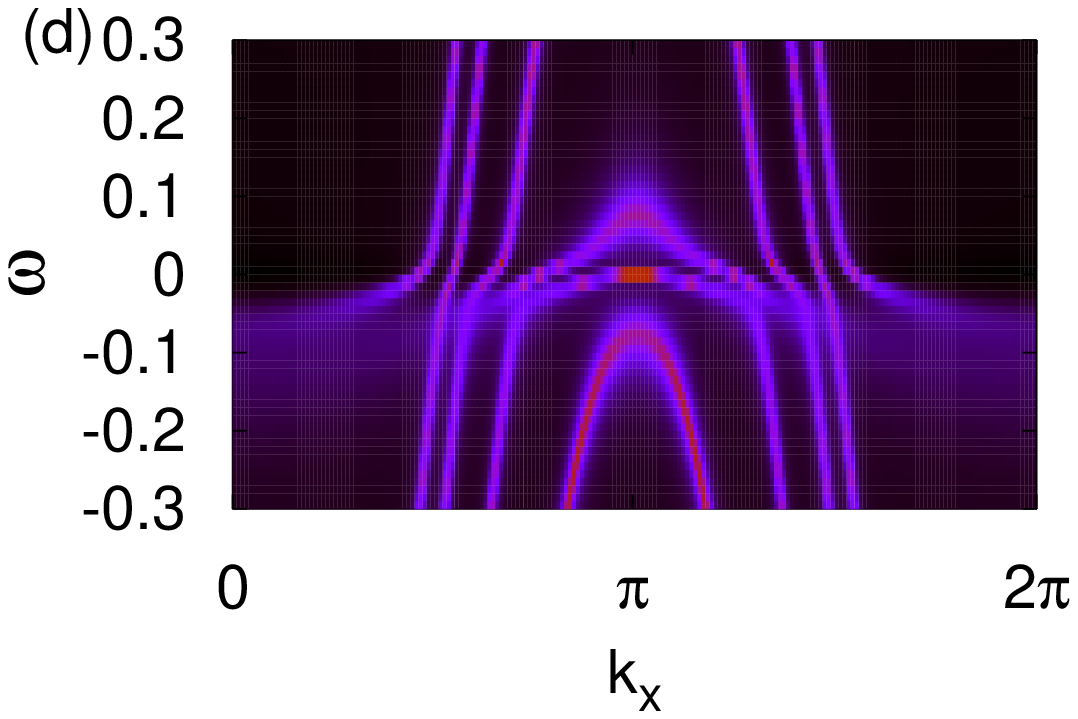}
\end{center}
\end{minipage}
\end{tabular}
\caption{(color online)
The spectral function $A(\omega,\vecc{k})$ at $(k_y,k_z)=(0,\pi/2L)$
for several temperatures when $(L_A,L_B)=(2,5)$. 
Temperatures for (a), (b), (c), and (d)
are $T=0.02, 0.015, 0.01, 0.0015$, respectively.
Violet corresponds to high-intensity regions and black corresponds to 
low-intensity regions.
}
\label{fig:Akx}
\end{figure}
\begin{figure}[htbp]
\begin{tabular}{cc}
\begin{minipage}{0.5\hsize}
\begin{center}
\includegraphics[width=\hsize,height=0.8\hsize]{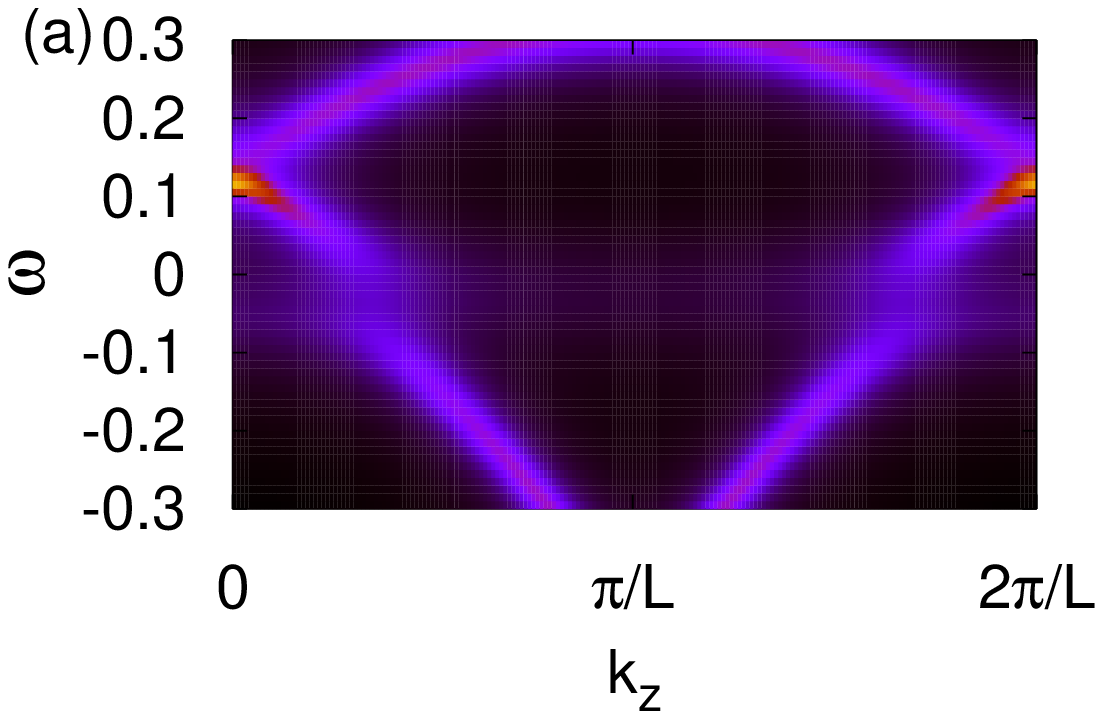}
\end{center}
\end{minipage}
\begin{minipage}{0.5\hsize}
\begin{center}
\includegraphics[width=\hsize,height=0.8\hsize]{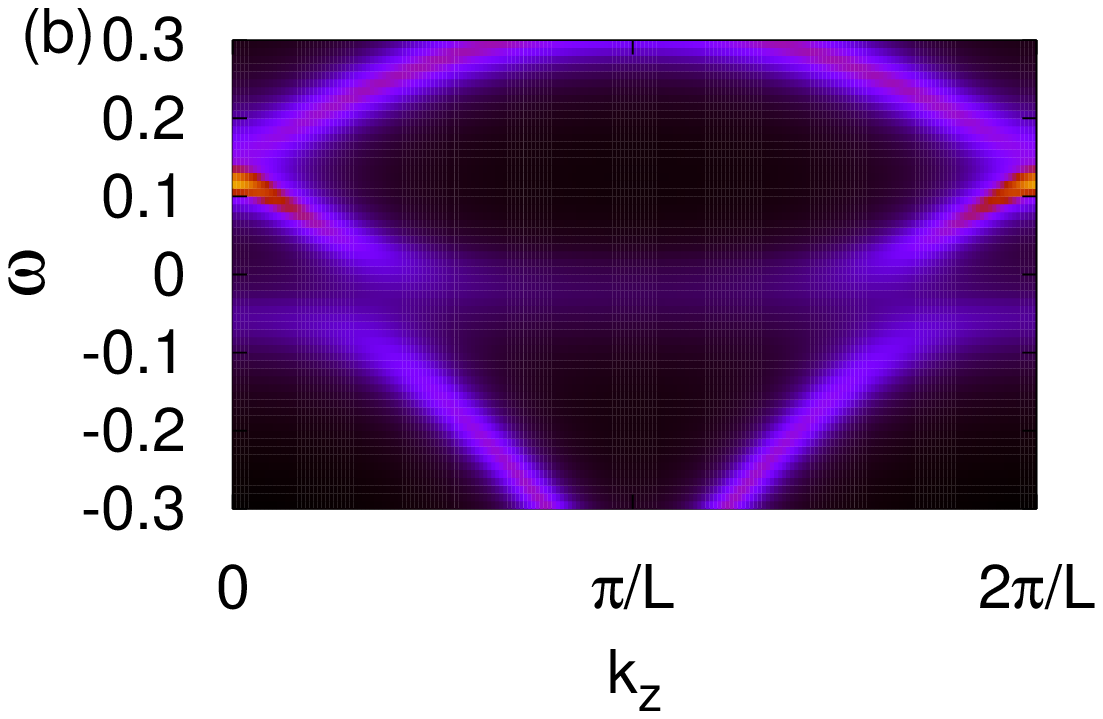}
\end{center}
\end{minipage}
\end{tabular}
\begin{tabular}{cc}
\begin{minipage}{0.5\hsize}
\begin{center}
\includegraphics[width=\hsize,height=0.8\hsize]{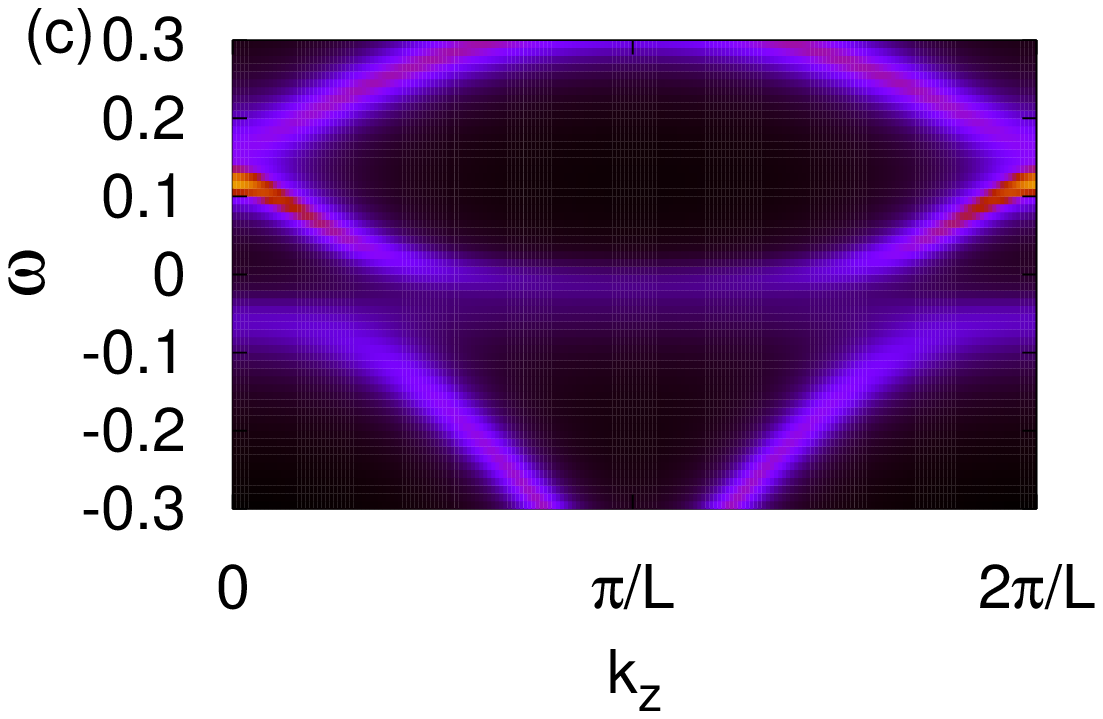}
\end{center}
\end{minipage}
\begin{minipage}{0.5\hsize}
\begin{center}
\includegraphics[width=\hsize,height=0.8\hsize]{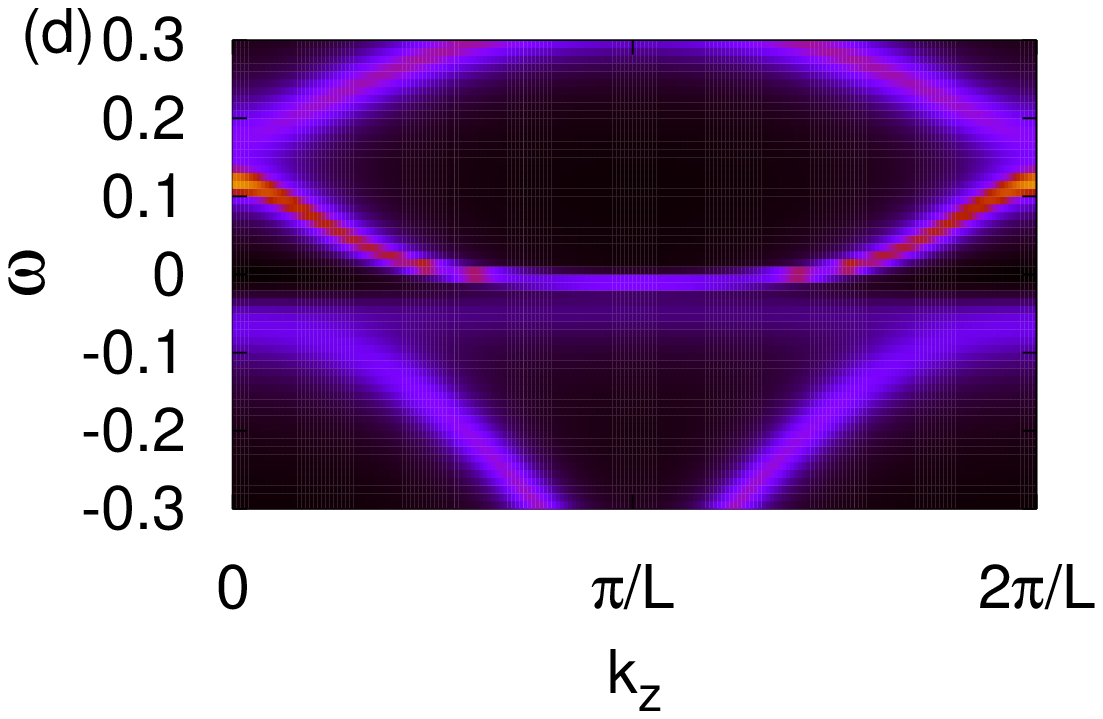}
\end{center}
\end{minipage}
\end{tabular}
\caption{(color online)
The spectral function $A(\omega,\vecc{k})$ at $(k_x,k_y)=(0.55\pi,0)$
for several temperatures when $(L_A,L_B)=(2,5)$. 
Temperatures for (a), (b), (c), and (d)
are $T=0.02, 0.015, 0.01, 0.0015$, respectively.
}
\label{fig:Akz}
\end{figure}
\begin{figure}[htbp]
\begin{tabular}{cc}
\begin{minipage}{0.5\hsize}
\begin{center}
\includegraphics[width=\hsize,height=0.8\hsize]{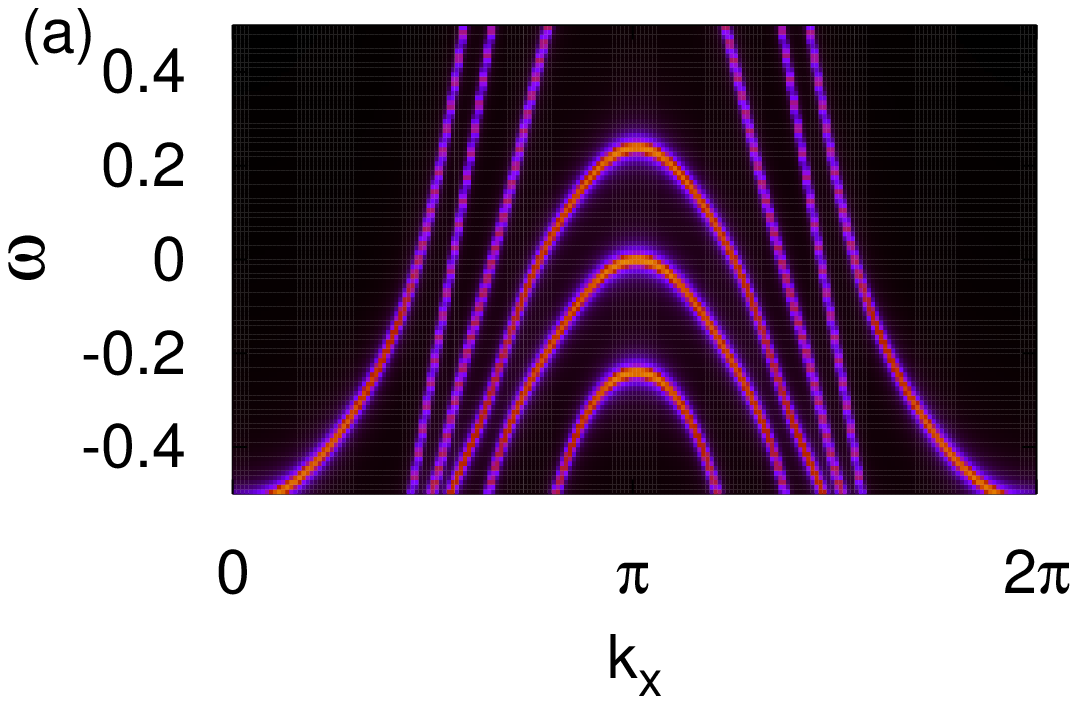}
\end{center}
\end{minipage}
\begin{minipage}{0.5\hsize}
\begin{center}
\includegraphics[width=\hsize,height=0.8\hsize]{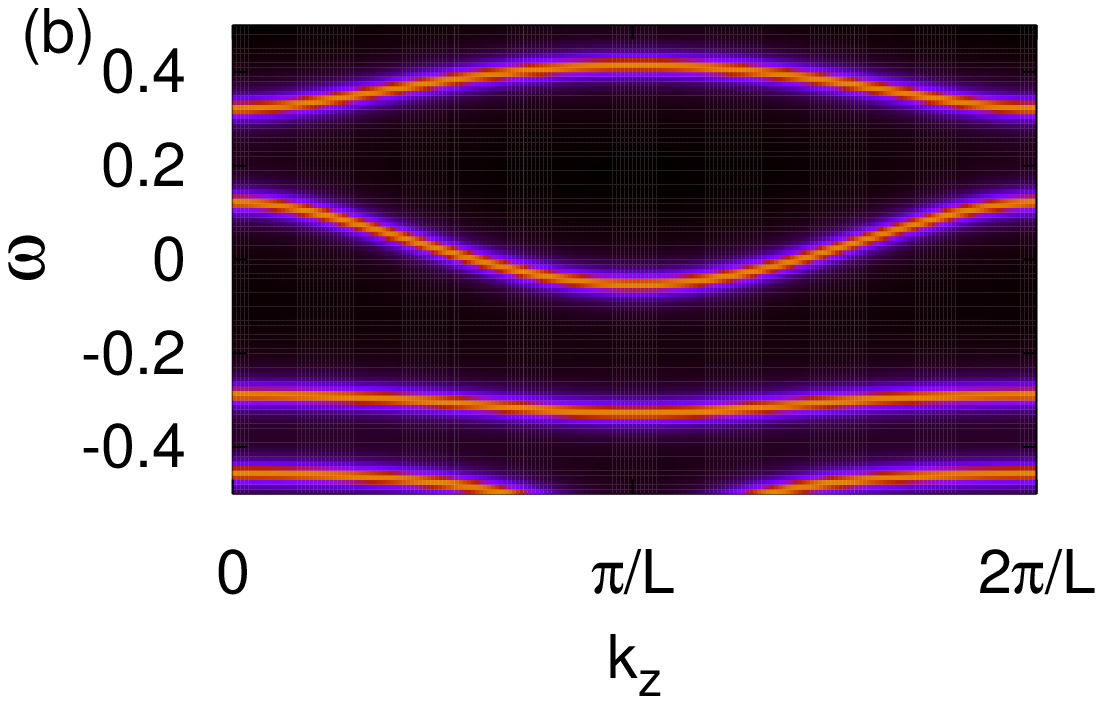}
\end{center}
\end{minipage}
\end{tabular}
\caption{(color online)
The spectral function $A(\omega,\vecc{k})$ for $U=0$
(a) at $(k_y,k_z)=(0,\pi/2)$ and (b) at $(k_x,k_y)=(0.55\pi,0)$
for $(L_A,L_B)=(2,5)$. 
}
\label{fig:Ak_U0}
\end{figure}

Indeed, finite curvatures along the $z$-direction
are seen in the dispersions of the superlattice.
Before discussing the dispersions along the $z$-direction,
we show contour plots of 
$A(\omega,\vecc{k})$ along the $x$-direction for several
temperatures at $(k_y,k_z)=(0,\pi/2L)$ 
for $(L_A,L_B)=(2,5)$ in Fig. \ref{fig:Akx}, as an example.
In the present study, 
a part of a band is called a heavy electron band,
if its broadening is small enough at low temperatures while it is
strongly smeared at high temperatures.
At $T=0.02$, there are 
no distinguishable
heavy electron bands around 
the Fermi energy $\omega=0$.
On the other hand, 
for $T=0.0015$, which is sufficiently lower than the crossover temperature
$T_0\sim 0.01$-$0.015$, 
the heavy electron bands are well formed around
$\omega=0$. 
Note that all the bands especially around $\omega\sim0$
are altered when the temperature
is changed in Fig. \ref{fig:Akx}, which means that the heavy electrons are
present in all bands at low temperatures.
Next, 
in Fig. \ref{fig:Akz}, we show the spectral function
along the
$z$-direction at $(k_x,k_y)=(0.55\pi,0)$
with the same parameters as in Fig. \ref{fig:Akx}.
The formation of the heavy electron bands is again observed 
around $T\sim 0.01$.
In order to distinguish the correlation effects in the spectral function,
dispersions for the non-interacting case, $U=0$, are also shown in
Fig. \ref{fig:Ak_U0} using the same parameters as in Figs.
\ref{fig:Akx} and \ref{fig:Akz}.
Interestingly, the correlation effects largely depend on the bands.
For example, in Fig. \ref{fig:Akx} (d), the most outer band at
$k_x=0\sim \pi/2$ is strongly renormalized and broadened compared to
other bands.
Similarly, in Fig. \ref{fig:Akz} (d), the nearly flat band around $\omega\sim
-0.05$ has weaker intensity than those of 
other bands around $|\omega|\sim 0.05$.
In the present superlattice structure, $(L_A,L_B)=(2,5)$, 
although there is only one kind of the local selfenergy
$\Sigma_{z=L\tilde{z}_1}=\Sigma_{z=L\tilde{z}_1+1}$, 
such band dependence in the spectral function 
can indeed arise due to the superlattice structure.

As already mentioned above,
the heavy electron bands have finite curvature along
the $z$-direction.
The width of the heavy electron bands
along the $z$-direction in Fig. \ref{fig:Akz}
is not negligiblly small compared to that along the $x$-direction in
Fig. \ref{fig:Akx}.
This is the typical behavior for general $(k_x,k_y)$ around the Fermi surface.
If the $c$-electrons in the B-layers 
do not participate in the formation of the heavy electrons,
heavy bands with such a large curvature for the $z$-direction
cannot be observed.
Indeed, 
the large curvature in the heavy bands is mainly due to the 
$c$-electron hopping which connects separated A-layers.
Therefore, the heavy electrons 
are present 
with significant weight
even in the B-layers,
and  the corresponding 
heavy electron wave functions are extended over the entire system \cite{pap:Peters2013}.
This is in strong contrast to the implicit assumption
in the previous studies~\cite{pap:Goh2012,pap:She2012}
that the heavy electrons exist only in the Ce-layers.
We emphasize that the heavy electron bands become observable 
at the same temperature
$T_0$ both for the $x$-direction and the $z$-direction,
which means that
there is no distinguishable 2D-3D dimensional crossover in the dispersion.
In our system, it is expected that any small $V>0$ makes
the $f$-electrons 3-dimensional at $T\ll T_0$ ($T_0$ depends on $V$),
while
they cannot coherently move in any direction
at $T\gg T_0$.
In this sense, looking only at the dispersions, 
there is no 2-dimensional temperature region in our system with 
$(L_A,L_B)=(2,5)$.
For other superlattice structures $(L_A,L_B)=(4,1), (3,1), (2,1), (2,2)$ and 
$(2,3)$ analyzed in the present study, 
the dispersions are qualitatively the same as that for 
$(L_A,L_B)=(2,5)$, although effect of the interaction becomes smaller
as a ``$f$-electron layer density" $L_A/L$ is decreased.

However, the formation of the heavy electrons 
does not directly 
imply
metallic character of 
the system, which is experimentally defined through transport
properties. Theoretically, this is because
the former is a one-particle property while the latter is related to
two-particle correlations. 
We thus investigate the resistivity along the $x$ and $z$-directions
to clarify the metallic character of the superlattice, which
is a direct measure of the dimensionality of the electron motions.

We calculate the resistivity 
using the Kubo formula.
The conductivity in arbitrary units is given by
\begin{align}
\sigma_{\mu\mu}&=\lim_{\omega\rightarrow0}\frac{1}{i\omega}
[K^R_{\mu\mu}(\omega)-K^R_{\mu\mu}(0)],\\
K_{\mu\mu}(i\omega_n)&=
\int_0^{1/T}d\tau \langle T_{\tau}J_{\mu}(\tau)J_{\mu}(0)
\rangle e^{i\omega_n\tau},\\
J_{\mu}&=\sum_{k\sigma}\sum_{aa^{\prime}ll^{\prime}}A^{a\dagger}_{kl\sigma}
v^{aa^{\prime}}_{ll^{\prime}}(\vecc{k})
A^{a^{\prime}}_{kl^{\prime}\sigma},\\
v^{aa^{\prime}}_{ll^{\prime},\mu}(\vecc{k})&=
\frac{\partial H^{aa^{\prime}}_{ll^{\prime}}(\vecc{k})}{\partial k_{\mu}}.
\end{align}
The resistivity is simply calculated by $\rho_{\mu\mu}=1/\sigma_{\mu\mu}$.
In the present study, we neglect the vertex corrections in $K_{\mu\mu}$
because 
it is known that effects of the vertex corrections 
on the resistivity are small \cite{pap:Kontani2008,pap:Lin2009}.
Under this approximation,
$K_{\mu\mu}$ is evaluated as
\begin{align}
K_{\mu\mu}(i\omega_n)&=-\frac{T}{N}\sum_{aa^{\prime}=c,f}
\sum_{\varepsilon_mk,\{l_i\}}
v^{aa}_{l_1l_2\mu}v^{a^{\prime}a^{\prime}}_{l_3l_4\mu}\notag \\
&\qquad G_{l_3l_2}^{a^{\prime}a}(i\varepsilon_m,\vecc{k})
G_{l_1l_4}^{aa^{\prime}}(i\varepsilon_m+i\omega_n,\vecc{k}).
\label{eq:K}
\end{align}
After an analytic continuation, we have three terms proportional
to $G^RG^R, G^AG^A$ and $G^AG^R$, by using the retarded
(advanced) Green's functions $G^{R(A)}$ \cite{pap:Kontani2008}.
Since it is seen that 
the 
$G^{R(A)}G^{R(A)}$
terms 
are much smaller than 
the 
$G^AG^R$
term
in the present study, 
we safely neglect them in the numerical
calculations.

\begin{figure}[htbp]
\begin{tabular}{cc}
\begin{minipage}{0.5\hsize}
\begin{center}
\includegraphics[width=\hsize,height=0.8\hsize]{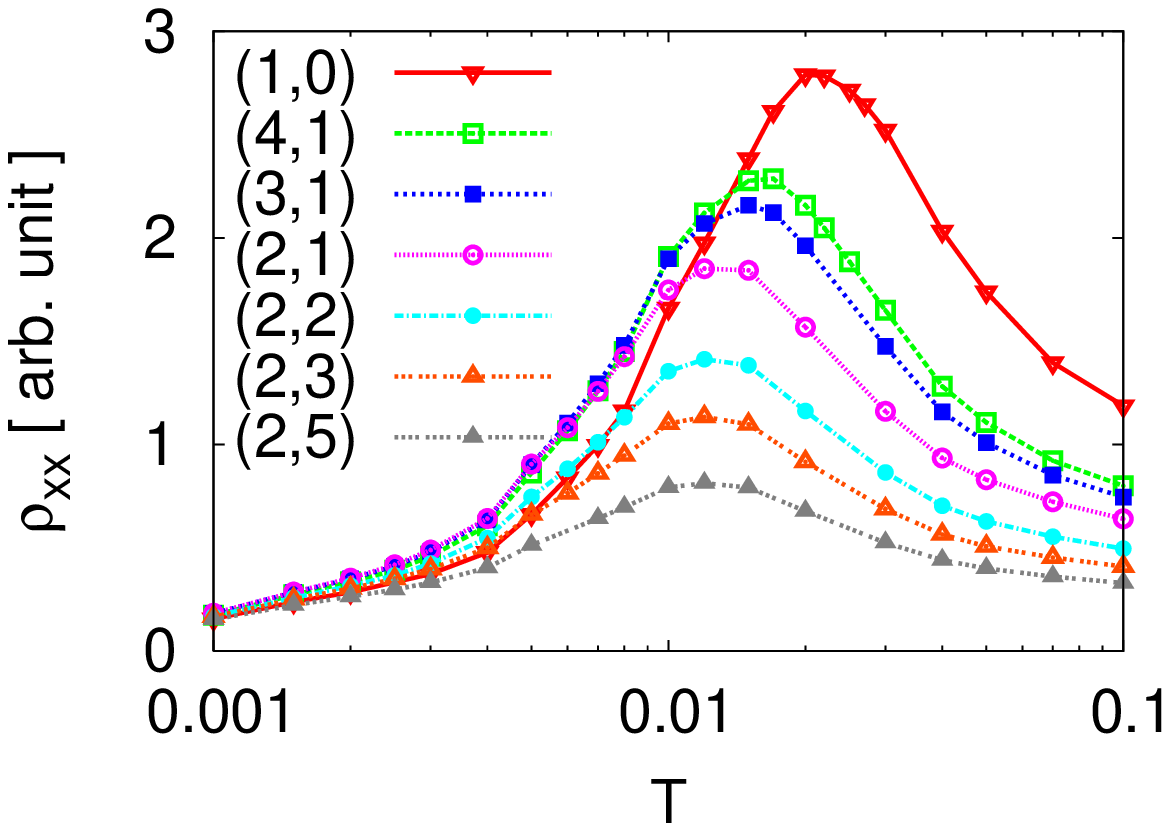}
\end{center}
\end{minipage}
\begin{minipage}{0.5\hsize}
\begin{center}
\includegraphics[width=\hsize,height=0.8\hsize]{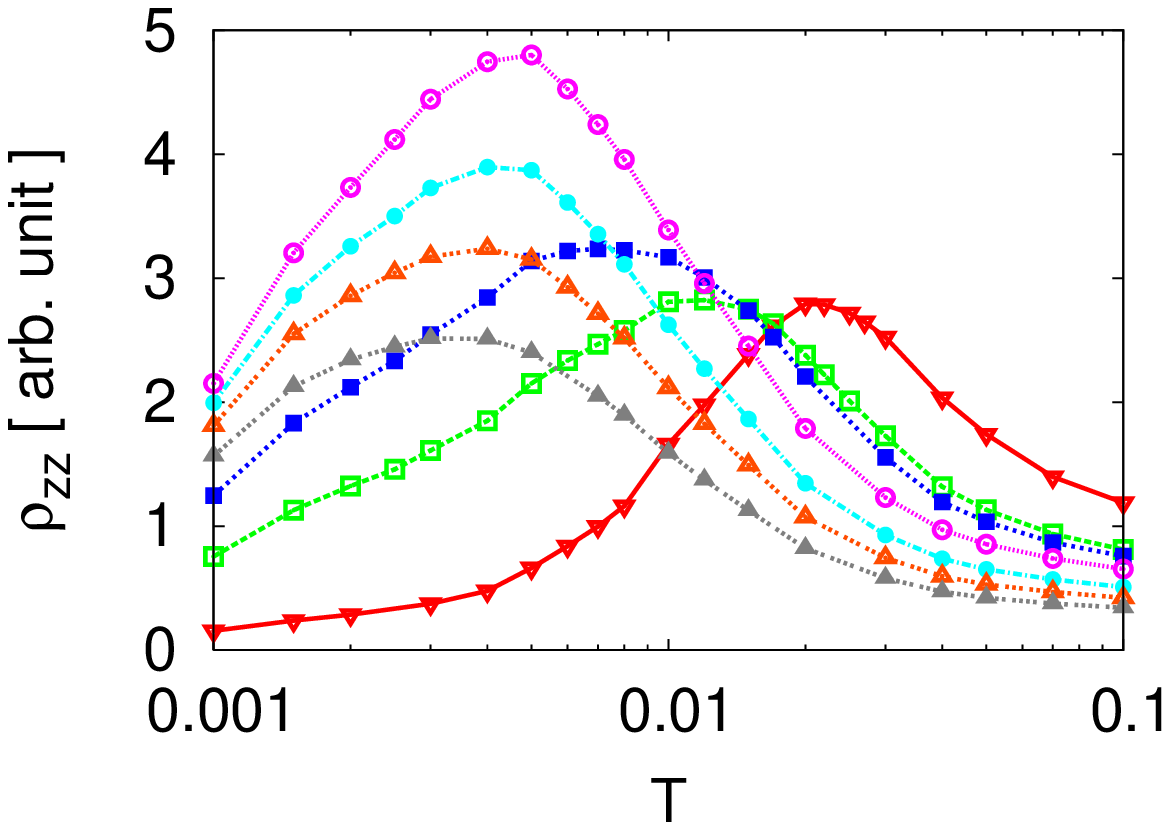}
\end{center}
\end{minipage}
\end{tabular}
\caption{(color online)
Temperature dependence of the resistivity for
the $x$-direction $\rho_{xx}$ (left panel) and the $z$-direction
$\rho_{zz}$ (right panel).
The numbers $(L_A,L_B)$ in the figure denote corresponding
layer configurations.
For a comparison, the resistivities for the 3D PAM $(L_A,L_B)=(1,0)$
is also shown with red curves.
}
\label{fig:s}
\end{figure}
In Fig. \ref{fig:s}, we show 
calculation 
results of the resistivities
in $x$-direction $\rho_{xx}$ and in $z$-direction
$\rho_{zz}$.
For a comparison, the resistivity for the bulk 3D PAM
$(L_A,L_B)=(1,0)$ is also shown as red curves.
The numbers in the figure $(L_A,L_B)$ represent layer configurations.
Similar to the 3D PAM, the resistivities in the superlattice exhibit peak
structures at the coherence temperatures 
below which the electron motions become coherent.
For decreasing the $f$-electron
layer density $L_A/L$,
the peak height of the in-plane resistivity $\rho_{xx}$ is suppressed and 
its positions $T_x$ are shifted to lower temperatures
compared with those of the bulk 3D PAM,
which would be
consistent with the experiments \cite{pap:Shishido2010,pap:Mizukami2012}.
In the limit $L_A/L\ll 1$, 
the conductivity becomes dominated by the $c$-electrons 
and effects of the interaction between 
the $f$-electrons get masked. 
The qualitative agreement with the experiments supports 
that our model calculations capture the essential physics
of $f$-electron superlattices.
If the superlattice is regarded
as a junction of a
light metal and a heavy metal with largely different Fermi velocities, 
as in the previous study \cite{pap:She2012},
the in-plane resistivity is supposed to be determined only by the light metal region
resulting in a monotonic temperature dependence~\cite{pap:Yamada1989}, 
which
  is in strong contrast to the experiments \cite{pap:Shishido2010,pap:Mizukami2012}.
Qualitatively similar $L_A/L$-dependence of the peak positions 
$T_z$ is also seen in the $z$-axis resistivity $\rho_{zz}$.
However, quantitatively, $\rho_{zz}$ much stronger depends on $L_A/L$.
Furthermore, the coherence temperature for the $z$-axis is lower than that for
the $x$-axis in the superlattice, and the
difference between both coherence temperatures grows as $L_A/L$ is reduced.

The difference between 
$T_x$ and $T_z$ can be 
explained, once 
the conductivity is divided into "intra-orbital"
and "inter-orbital" contributions,
$\sigma_{\mu\mu}=\sigma_{\mu\mu,{\rm intra}}+\sigma_{\mu\mu,{\rm inter}}$.
The former is defined by restricting $l_1=l_2=l_3=l_4$ in 
Eq. (\ref{eq:K}), and 
the latter is defined by a sum of all other terms.
If we define constituent currents by
$j_{\mu,ll^{\prime}}
\equiv \sum_{k \sigma} \sum_{aa'}
A^{a\dagger}_{k l \sigma} v^{a a'}_{ll',\mu}  A^{a'}_{kl'}$,
the separate contributions can be expressed in a simplified notation 
as $\sigma_{\rm intra}
\sim\sum_l \langle j_{ll}j_{ll}\rangle$ and
$\sigma_{\rm inter}
\sim\sum_{l_1,l_2,l_3,l_4}' 
\langle j_{l_1l_2}j_{l_3l_4}\rangle$, where $\sum_{l_1,l_2,l_3,l_4}'$
is defined as a summation over $l_1\sim l_4$ except for $l_1=l_2=l_3=l_4$.
The two contributions describe different transport processes, and each of them 
alone is not an observable.
We note that $\sigma_{\rm intra}$ must be non-negative,
since it is written by a sum of correlation functions of the constituent
currents $j_{ll}$ with itself.
On the other hand, $\sigma_{\rm inter}$ does not need to be non-negative,
since it is not written only by self-correlations of the constituent currents
and it includes other terms like $\langle j_{11}j_{23}\rangle$.

\begin{figure}[htbp]
\begin{tabular}{cc}
\begin{minipage}{0.5\hsize}
\begin{center}
\includegraphics[width=\hsize,height=0.8\hsize]{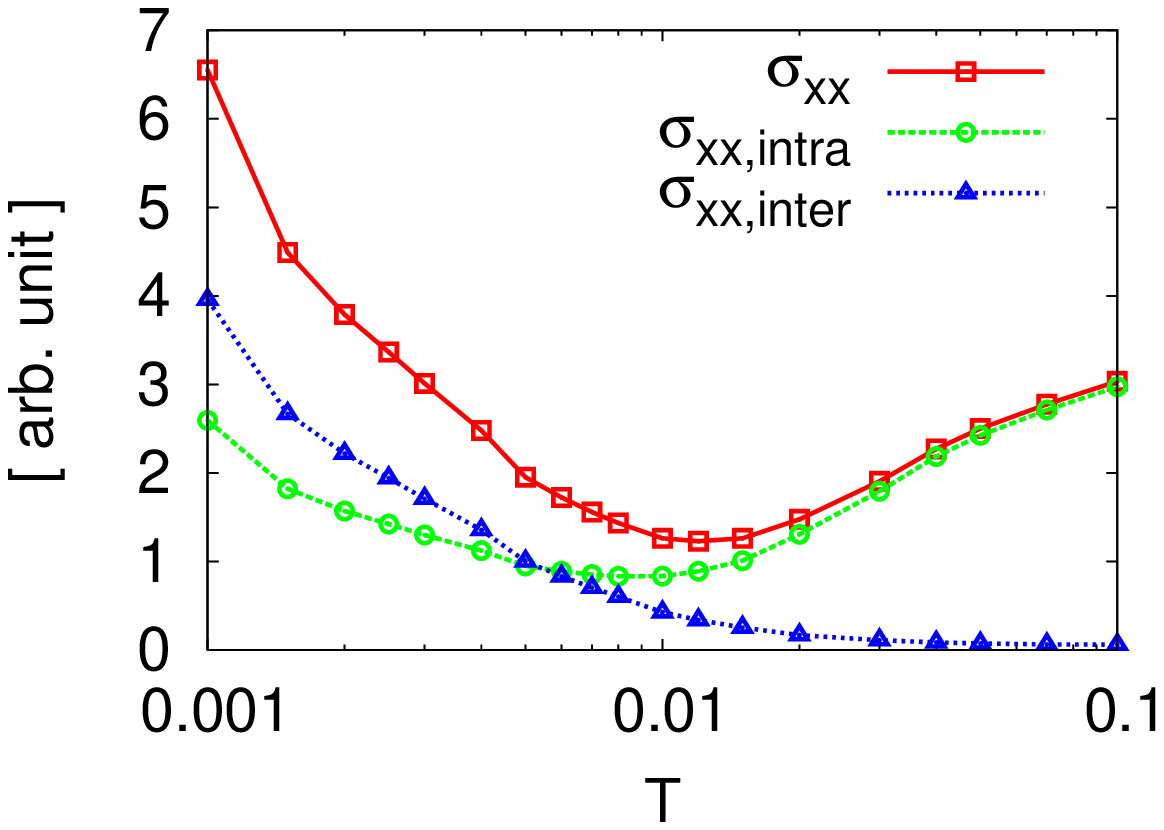}
\end{center}
\end{minipage}
\begin{minipage}{0.5\hsize}
\begin{center}
\includegraphics[width=\hsize,height=0.8\hsize]{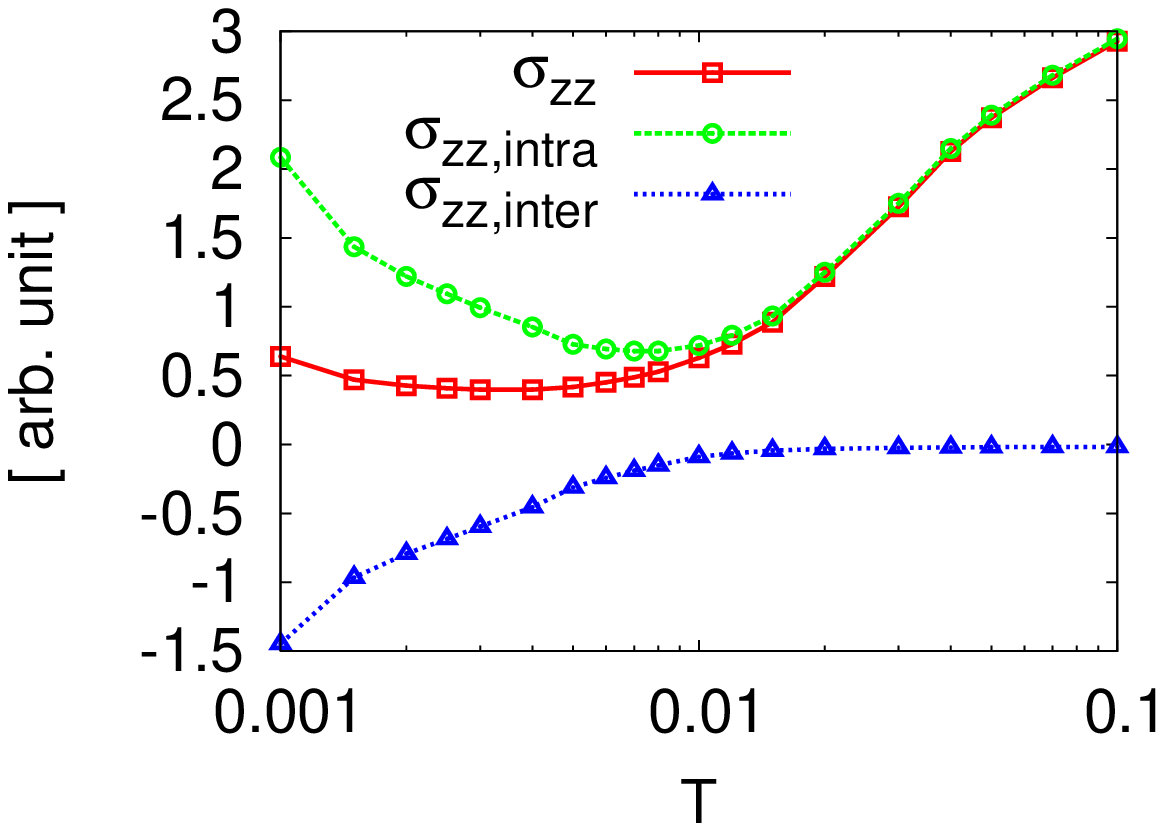}
\end{center}
\end{minipage}
\end{tabular}
\caption{(color online)
Temperature dependence of the $x$-axis conductivity (left panel)
and $z$-axis conductivity (right panel) for $(L_A,L_B)=(2,5)$.
}
\label{fig:s_int}
\end{figure}
Figure \ref{fig:s_int} 
shows $\sigma_{\mu\mu}$, $\sigma_{\mu\mu,{\rm intra}}$ and $\sigma_{\mu\mu,{\rm inter}}$
for $\mu=x,z$ when $(L_A,L_B)=(2,5)$ as an example.
At high temperature $T>T_x$, the conductivities are determined by
the intra-orbital contributions. We note that 
$\sigma_{xx,{\rm intra}}$ and $\sigma_{zz,{\rm intra}}$ 
exhibit a minimum at the same
temperature scale $T_x\simeq T_0$ below which the heavy electrons are
well-defined. On the other hand, 
at low temperatures $T<T_x$, the inter-orbital contributions become
important. While $\sigma_{xx,{\rm inter}}$ is positive, 
$\sigma_{zz,{\rm inter}}$ is 
negative
at low temperatures, so that it strongly suppresses $\sigma_{zz}$
for the present parameters.
The negative contribution to the total conductivity means that the transport processes corresponding to $\sigma_{zz,{\rm inter}}$ increase resistivity.
Since $\sigma_{\mu\mu,{\rm inter}}$ is mainly determined by 
$G_{ll^{\prime}}(l\neq l^{\prime})$, 
we conclude 
that
the reduction of the $z$-axis conductivity is due to scattering between different RBZs.
Such transport processes 
give positive contributions to
$\sigma_{xx}$, because the $k_z$ dependence is
not important for the $x$-direction transport.
Namely, the heavy electrons are scattered
by the superlattice structures along the $z$-axis resulting in
the reduced conductivity $\sigma_{zz}$, while such scattering does not
affect the in-plane conductivity $\sigma_{xx}$.

\begin{figure}[htbp]
\includegraphics[width=0.8\hsize]{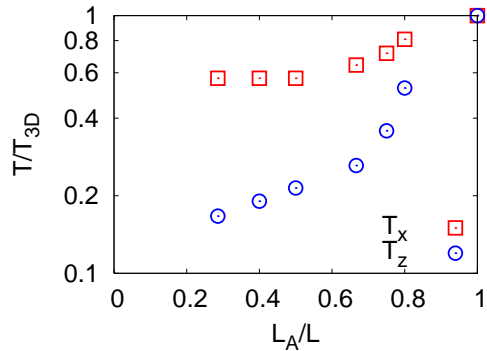}
\caption{(color online)
Coherence temperatures vs $L_A/L$
for $(L_A,L_B)=(1,0), (4,1), (3,1), (2,1), (2,2), (2,3)$, and (2,5).
}
\label{fig:phase}
\end{figure}
We now summarize 
the dependence of the coherence
  temperatures $T_{x,z}$ 
on the $f$-electron layer density $L_A/L$ in Fig. \ref{fig:phase}.
These temperatures 
correspond to the
energy scales for the coherent motion of the
heavy electrons in different direction ($x$- and $z$-direction).  
From those, we can draw conclusions about the 
dimensionality of the heavy electron motions.
The $L_A/L$-$T$ phase diagram has three distinct regions:
a high temperature region 
($T>T_x$), a $2$D-like region ($T_z<T<T_x$),
and an anisotropic $3$D-like region ($T<T_z$).
In the high temperature region, the heavy electrons are not
well-defined. 
On the other hand, when $T<T_z$, the system is an anisotropic 3D
Fermi liquid metal 
and the heavy electrons
can move coherently in any direction. 
The most 
remarkable
region, however,  is the $2$D-like region where coherence 
is only well developed in the $xy$-plane,
while for the $z$-direction coherent movement 
is strongly suppressed.
In this region, the heavy electron motions are 
two dimensional. 
As the temperature is lowered for fixed $L_A/L$, or $L_A/L$ is reduced for
fixed temperature, 
a crossover from two to three dimensions in the behavior
  of the heavy electrons takes place.
In the limit $L_A/L\rightarrow 0$ 
the resistivities are completely dominated  by
the $c$-electrons and the peaks are smeared, which suggests the 
system behaves nearly as 
a $3$D free electron system and there is no longer a clear dimensional
crossover.

We emphasize that the dimensional crossover is intrinsic in the superlattice. 
This can be compared to
layered systems without superlattice
structures. 
While $\sigma_{xx}$ and $\sigma_{zz}$ differ in
  magnitude in such systems, 
their temperature dependence is essentially the same and $T_{x}, T_z$ 
are supposed to coincide~\cite{pap:Valla2002}.
This holds true,
if the momentum dependence of the selfenergy is 
weak,
as in a case of formation of the canonical Fermi liquid
\cite{book:Yamada2004}.
Our results suggest that 
the superlattice structure is 
important
 for 
observing the
dimensional crossover.
As far as we know, the present study is the first demonstration
of dimensional crossover in $f$-electron systems. 
We have revealed the nature of the heavy electrons,
 which is essential
in understanding many physical properties,
in layered $f$-electron superlattices.
While most of the present analyses are directly applicable only to
the paramagnetic states, they serve as a basis for understanding
the intriguing experiments on the ordered states.

\section{discussion and summary}
Finally, we give a qualitative discussion  on the
  experiments, based on the dimensional crossover 
as a possible explanation.
For this discussion, we have to keep in mind that $T_x$ and $T_z$ 
are defined by the resistivities in the normal paramagnetic state,
and both of the $c,f$-electron contributions are important in the transport
while $f$-electron contributions would be crucial in itinerant magnetism and
superconductivity.
However, the dimensionality in the transport which is a direct measure of 
the metallic character of the system
could be closely
related to the $z$-axis coherence length 
of the experimentally observed ordered states. 
When $L_A/L$ is small, magnetic coupling between separated Ce-layers 
is supposed to be suppressed. This leads to low N\'eel temperatures,
which is consistent
with
the experiments in CeIn$_3$/LaIn$_3$ \cite{pap:Shishido2010}.
In such a case, spin fluctuations can exhibit 2D character
in some temperature regions~\cite{pap:Kondo2002,pap:Garst2008}.
Furthermore, the dimensionality of the heavy electrons would also be relevant
for $H_{c2}$ of the superconductivity in CeCoIn$_5$/YbCoIn$_5$
when the orbital-depairing is dominant.
For small $L_A/L$, 2D-like movement of the heavy electrons would
result in strong field angle dependence of $H_{c2}$ around the
zero-field transition temperature $T_{c0}$ as found
in the experiments~\cite{pap:Mizukami2012,pap:Goh2012}.
We point out that this is in sharp contrast to 
the previously studied normal-metal-superconductor
superlattices such as Ni/Cu and V/Ag 
\cite{pap:Klemm1975,pap:Takahashi1986,
pap:Barnerjee1983,pap:Kanoda1986,pap:Jin1989} where
the anomalous angle dependence of $H_{c2}$ is seen for sufficiently lower
temperatures than $T_{c0}$.
Although the present results 
and the experiments have common tendencies 
concerning dimensionality and temperature dependence,
the discussions of the experiments presented here are merely qualitative.
Itinerant magnetic properties and superconducting properties cannot be
described within the DMFT+NRG which is used in the present study.
Detailed investigations of these properties 
are left for future studies.

In summary, we have investigated the
$f$-electron layered superlattice within DMFT+NRG.
The heavy electrons are formed in the entire system below $T_0$ as seen
in the spectral function.
However, we have identified in the resistivity two
  distinct energy scales for 
  the coherent motion of 
the heavy electrons satisfying $T_z<T_x\simeq T_0$.
The 
results of $\rho_{xx}$ are qualitatively consistent
with the experiments, which supports our model calculations.
We find that the heavy electron motions show a dimensional crossover
between two and three dimensional character.
This dimensional crossover would be responsible for 
the behaviors of the AF and SC in the 
CeIn$_3$/LaIn$_3$ and CeCoIn$_5$/YbCoIn$_5$ superlattices.
Our present results thus build the  basis for
understanding the $f$-electron superlattice and the related experiments.

We thank
Y. Matsuda, T. Shibauchi, H. Shishido, H. Ikeda, H. Kusunose,
K. Ueda, S. Fujimoto, and N. Kawakami for valuable discussions. 
This work is supported by JSPS/MEXT KAKENHI
Grant Numbers 23840009 (Y.~T.) and 20102008 (M.~O.).
RP thanks the Japan Society for the Promotion of Science (JSPS) for
the support by its FIRST Program.

\bibliography{sulattice}

\begin{thebibliography}{29}
\bibitem[{com()}]{com:NRG}
\bibinfo{note}{We use $\Lambda=2$ and $N_{\rm kept}=1200$ kept states in the
  NRG-calculations.}
\end{thebibliography}

\end{document}